\pdfoutput=1
\documentclass[aps, prb, amsfonts, amsmath, twocolumn, superscriptaddress, 10pt, floatfix]{revtex4-1}
\usepackage{natbib}

\usepackage[colorlinks=true]{hyperref}
\usepackage{graphicx}
\usepackage{amsmath}
\usepackage{units}
\usepackage{color}
\usepackage[caption=false]{subfig}

\usepackage{amssymb}

\usepackage{bm}										% bold in math
\usepackage{braket}
\usepackage{mathtools}								% to put text on top of arrows

\newcommand{\mathif}[1]{\relax\ifmmode#1\else$#1$\fi}

\newcommand{\TT}{\mathif{\text{TT}}}
\newcommand{\lep}{\mathif{\text{LE}^+}}
\newcommand{\lem}{\mathif{\text{LE}^-}}
\newcommand{\lepm}{\mathif{\text{LE}^\pm}}
\newcommand{\ctp}{\mathif{\text{CT}^+}}
\newcommand{\ctm}{\mathif{\text{CT}^-}}
\newcommand{\ctpm}{\mathif{\text{CT}^\pm}}

\newcommand{\sm}{Supplementary}
\newcommand{\norm}[1]{\left\lVert#1\right\rVert}
\begin{document}
\graphicspath{{./}{./img/}}
	
\title{ Multi-dimensional Tensor Network Simulation of Open Quantum Dynamics in Singlet Fission }

\author{Florian A. Y. N. Schr\"oder}
\affiliation{Cavendish Laboratory, University of Cambridge, J. J. Thomson Avenue, Cambridge CB3 0HE, United Kingdom}
\author{David H. P. Turban}
\affiliation{Cavendish Laboratory, University of Cambridge, J. J. Thomson Avenue, Cambridge CB3 0HE, United Kingdom}
\author{Andrew J. Musser}
\affiliation{Department of Physics and Astronomy, University of Sheffield, Hounsfield Road, Sheffield S3 7RH, United Kingdom}
\author{Nicholas D. M. Hine}
\affiliation{Department of Physics, University of Warwick, Gibbet Hill Road, Coventry CV4 7AL, United Kingdom}
\author{Alex W. Chin}
\email[Corresponding Author:]{ac307@cam.ac.uk}
\affiliation{Cavendish Laboratory, University of Cambridge, J. J. Thomson Avenue, Cambridge CB3 0HE, United Kingdom}

\begin{abstract}% 145 words, max 150
We develop a powerful tree tensor network states method that is capable of simulating exciton-phonon quantum dynamics of larger molecular complexes and open quantum systems with multiple bosonic environments.
We interface this method with \emph{ab initio} density functional theory to study singlet exciton fission (SF) in a pentacene dimer.
With access to the full vibronic many-body wave function, we track and assign the contributions of different symmetry classes of vibrations to SF and derive energy surfaces, enabling us to dissect,
understand, and describe the strongly coupled electronic and vibrational dynamics, relaxation, and reduced state cooling.
This directly exposes the rich possibilities of exploiting the functional interplay of molecular symmetry, electronic structure and vibrational dynamics in SF material design.
The described method can be similarly applied to other complex (bio-) molecular systems, characterised by a rich manifold of electronic states and vibronic coupling driving non-adiabatic dynamics.
\end{abstract}

\maketitle
Ultrafast open quantum dynamics have recently attracted considerable attention in the context of molecular light harvesting materials, where experimental evidence has begun to highlight the role of coherent, 
non-equilibrium vibronic dynamics in systems ranging from organic photovoltaics to the pigment-protein complexes of photosynthesis \cite{bredas2017photovoltaic,romero2017quantum}. 
Exploiting this complex interplay of vibrational and electronic quantum dynamics in tailored materials offers an exciting route towards advanced, next-generation light harvesting devices \cite{scholes2017using}, 
and the process known as singlet fission (SF) has recently been shown to be a very promising area in which to explore this. 

In organic semiconductors such as pentacene, a photogenerated singlet exciton undergoes singlet fission into a pair of entangled triplet excitons on sub-$\unit[100]{fs}$ timescales, 
efficiently generating two electronic excitations from a single incoming photon\cite{Smith2013,Berkelbach2013}.
Harnessing this potential carrier multiplication could help to overcome the Shockley-Queisser limit in photovoltaics \cite{Shockley1961}, 
and understanding the links between the dynamics and efficiency of SF has emerged as an active interdisciplinary field of research \cite{Wilson2013, Walker2013a,Yost2014,Alguire2015,Tamura2015,Zheng2016}.
Recently, ultrafast optical spectroscopy experiments have revealed how non-equilibrium and non-perturbative open quantum dynamics contribute to the kinetics of SF,
highlighting the role of the molecular vibrational environment in driving ultrafast SF through rapid energy relaxation, conical intersections and vibronic mixing effects \cite{Bakulin2015,musser2015evidence,Fuemmeler2016,Miyata2017,Stern2017,tempelaar2017vibronic}. 
Importantly, these results indicate that the \emph{evolving}, high-dimensional quantum states of the environment can become strongly \emph{correlated} with the fate of electronic photoexcitations.
As in many of the examples mentioned above, this entanglement leads to physics that can only be explored with numerical techniques moving substantially beyond the Born-Markov master equations that treat the environment as a simple `heat bath' \cite{Berkelbach2013b,Morrison2017}. 

However, obtaining accurate real-time information about the environmental quantum state, which is essential for determining and controlling complex reaction pathways, is a major theoretical challenge, 
as it requires the simulation of exponentially large vibronic wave functions.
While a powerful and popular approach to this in chemical physics is the Multi-Configurational Time-Dependent Hartree-Fock (MCTDHF) algorithm \cite{wang2003multilayer,Worth2004,Zheng2016},
we develop upon the deep insights into many-body wave function compression exploited in the highly efficient Matrix Product State (MPS) and Tensor Network State (TNS) ans\"atze which are widely used in correlated condensed matter problems \cite{Schollwoeck2011,Orus2014,Vidal2007}.
Encouragingly, MPS ans\"atze have already been shown to provide highly accurate results including hundreds of quantised and highly excited vibrations for model open systems \cite{Chin2010,chin2013role,Guo2012,prior2010efficient,Schroder2016,Wall2016},
but extensions to more realistic and higher-dimensional linear-vibronic models, as might be obtained from \textit{ab initio} electronic structure simulations, have not yet been possible.

\begin{figure}
	\includegraphics[width=\linewidth]{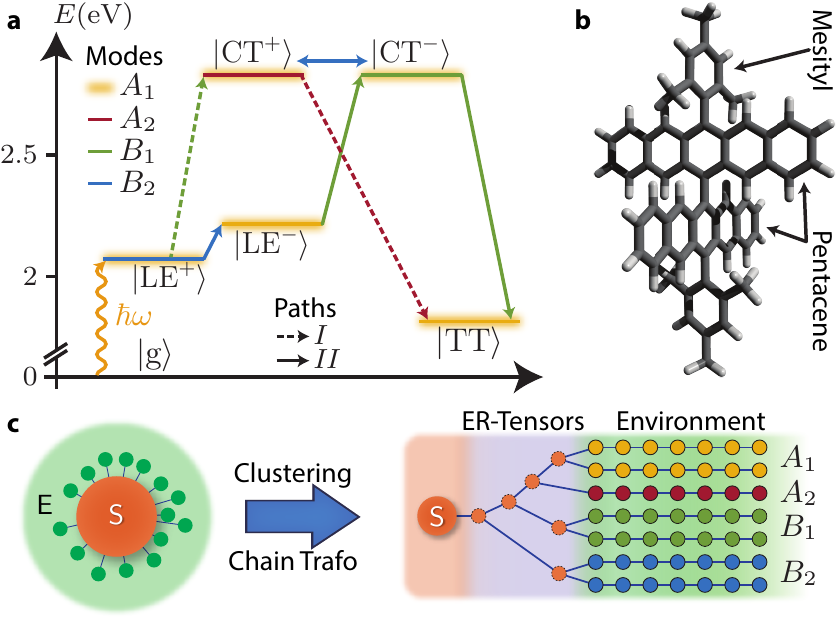}
    \caption{\textbf{Reaction pathways and modelling.}
    	(a) Electronic states and their coupling through vibrational modes in DP-Mes (b), coloured by their symmetry within the point group $D_{2d}$.
        After photoexcitation of $\lep$, SF can proceed via pathway I (dotted) or II (lines).
        (c) Once clustering identified symmetries and coupling patterns within the full vibrational environment (E), each cluster of modes is transformed onto chains to yield the model Hamiltonian.
        In the TTNS (right), all environmental states are connected to the electronic states (S) through a network of entanglement renormalisation (ER) tensors to improve speed and accuracy.
	}
	\label{fig:states}
\end{figure}

In this article, we present a technique based on a Tree Tensor Network State (TTNS) representation of the many-body vibronic wave function that exploits recent advances in MPS theory \cite{Schroder2016,Haegeman2014}, machine learning, 
and entanglement renormalisation methods\cite{Vidal2007} to provide such a capability. 
We demonstrate our suite of methods on an \textit{ab initio} DFT-parameterised model of 13,13'-bis(mesityl)-6,6'-dipentacenyl (DP-Mes, Fig.~\ref{fig:states}b), a pentacene-based dimer molecule in which SF has been observed on ultrafast (sub-ps) timescales,
despite being symmetry forbidden at its ground state geometry \cite{Lukman2015,Lukman2016}. 
Within a vibronic Hilbert space of $10^{500}$ states, we show how SF arises from a interconnected \emph{sequence} of symmetry breaking motions, directly pinpointing - through environmental observables -
how ultrafast reorganisation of \emph{distinct} environmental modes spontaneously \emph{create} electronic couplings that give $\approx 90\%$ TT yield. 
Even better, we also show that the formalism of TTNS naturally provides a completely general and intuitive `on the fly' method to discover the dominant many-body configurations of the environment in the dynamics,
allowing us to visualise non-equilibrium processes on environmental energy surfaces. 
With this new capability, we unambiguously show that SF occurs in DP-Mes via an avoided crossing, rather than a conical intersection,
highlighting the potential utility of our technique for unravelling the complex dynamics in a wide range of open quantum systems across physics and chemistry.

\textbf{Electronic structure and vibronic model.}
% system Hamiltonian
\textit{Ab initio} TD-DFT calculations have been performed on DP-Mes to derive a microscopic Hamiltonian model, spanned by five electronic diabatic states, denoted system, directly relevant to SF at the novel `cruciform' geometry of its ground state (see Fig.~\ref{fig:states} and \sm{} Fig.~1).
These states are the (anti)symmetrised local exciton states $\lepm$, the (anti)symmetrised charge-transfer states $\ctpm$ and the final triplet pair TT.
The $\lepm$ and $\ctpm$ states are coherently delocalised over the two monomer units, leading to an optically bright $\lep$ and dark $\lem$ state (J-dimer).
We find excited state energies of $\unit[2.07]{eV}$ ($\lep$), $\unit[2.20]{eV}$ ($\lem$), $\unit[2.75]{eV}$ ($\ctpm$) and $\unit[1.83]{eV}$ (TT), which define the diagonal of the system Hamiltonian $\bm H_S$.
This makes for an ideal system for studying the role of vibronic quantum dynamics in SF, as, crucially, the molecular symmetry of the ground state geometry suppresses \emph{all} electronic coupling between excited states.
Thus, although SF is strongly exergonic in this dimer it is strictly forbidden unless this symmetry is broken by vibrational motion (see \sm{} Fig.~2).
However, as in many SF systems, even symmetry breaking cannot induce direct coupling between $\lep$ and TT.
Instead, this is indirectly mediated by the high-lying CT states via super-exchange \cite{Smith2013, Berkelbach2013, Berkelbach2013b}, which is why they must be included in both the \textit{ab initio} and dynamical simulations, even if they are never populated \cite{Berkelbach2013}.

% Linear vibronic coupling model
We then employ further DFT calculations to construct a linear vibronic Hamiltonian\cite{Kuppel1984}
\begin{equation}
	\hat{\bm H}_{LVC} = \bm H_{S} + \sum_{n=1}^{252} \bm W_n \frac{(\hat b^\dagger_{n} + \hat b_{n})}{\sqrt2}+\hat H_{vib},
\end{equation}
based on $252$ quantum harmonic normal modes $\hat b_n^\dagger$ of the dimer's ground state potential.
This describes the emerging electronic couplings $\bm W_n$ as the system is perturbed from the ground state geometry (see \sm{} Sec.~II).

\textbf{Interface to TTNS.}
% Clustering and Hamiltonian transformation
Following previous t-DMRG and Matrix Product State (MPS) approaches, we then perform a unitary transformation to cast $\hat H_{vib}$ into an equivalent system of independent 1D-chain environments (Fig.\ref{fig:states}c).
The orthogonal polynomial chain transformation\cite{Chin2010,prior2010efficient,Schroder2016} directly exposes the low entanglement properties of the environments, which are essential for its efficient simulation with tensor wave functions (\textit{vide infra}).
However, the transformation can only be applied to modes with identical coupling pattern $\overline{\bm W}_n$ and arbitrary strengths $\lambda_n$, where $\bm W_n =\lambda_n \overline{\bm W}_n$.
While this is often the case in toy models (allowing the entire environment to be lumped into a single chain), our \textit{ab initio} patterns $\overline{\bm W}_n$ show $n$-dependent variations that grow with the irregularity and size of the molecular structure.
In order to group modes with similar $\overline{\bm W}_n$, we employ unsupervised machine learning in the form of k-means clustering\cite{Arthur2007}.
Thereby, we can find the minimal required set of environments $i$ and their coupling operators $\overline{\bm W}_i$ that optimally approximate $\overline{\bm W}_n$ of assigned modes.
We find that DP-Mes couplings require a minimum of four clusters, as confirmed by group theory, one for each 1D irreducible representation (irreps) $A_{1/2}$ and $B_{1/2}$ of the approximate point group $D_{2d}$ of DP-Mes.
However, the large variances of the couplings are only sufficiently described by seven clusters (see \sm{} Fig.~3).

The transformation of each cluster into chains $\hat H_{c,i}$ yields the star-like Hamiltonian (see Fig.~\ref{fig:states}c)
\begin{equation}
	\hat{\bm H}_{star} = \bm H_{S} + \sum_{i=1}^7 \left[ \overline{\bm W}_i \norm{\bm \lambda_{i}} \frac{(\hat a^\dagger_{i,0} + \hat a_{i,0})}{\sqrt2}+ \hat H_{c,i} \right],
\end{equation}
where $\bm \lambda_i = (\lambda_{i_1},\ldots,\lambda_{i_n})$, and $a_{i,0}^\dagger$ is termed the reaction coordinate (RC) of chain $i$, which directly reflects collective motion and time scales of the bath.
The matrices $\overline{\bm W}_i$ and parameters of our model are given in the \sm{} Sec.~II.

\textbf{TTNS for vibronic quantum dynamics.}
% TTNS from MPS, ref to reviews
The model is simulated with tree tensor network states (TTNS) \cite{Shi2006,Szalay2015}, which extend matrix-product state (MPS) methods for open quantum systems\cite{Schollwoeck2011,Verstraete2004,prior2010efficient,Guo2012,Orus2014,Schroder2016,Wall2016} 
to models with multiple bosonic environments.
%We briefly outline the key features of our TTNS, while for technical details on MPS we refer to the \sm{} Sec.~\ref{sec:si-ttns} and the excellent reviews given in Refs.~\onlinecite{Schollwoeck2011,Orus2014}. % TODO: remove this line?

The strength of this method lies in an optimal approximate compression of the full vibronic system-environment wave function.
It relies on the fact that the vast majority of the formally exponentially large Hilbert space is never explored, allowing a low-rank tensor approximation scheme that decomposes the wave function to yield a connected network of small,
individually updatable tensors that enable controllable approximation and efficient storage and manipulation of the full quantum state (see \sm{} Sec.~III and references within for formal details).

\begin{figure}
	\includegraphics[width=\linewidth]{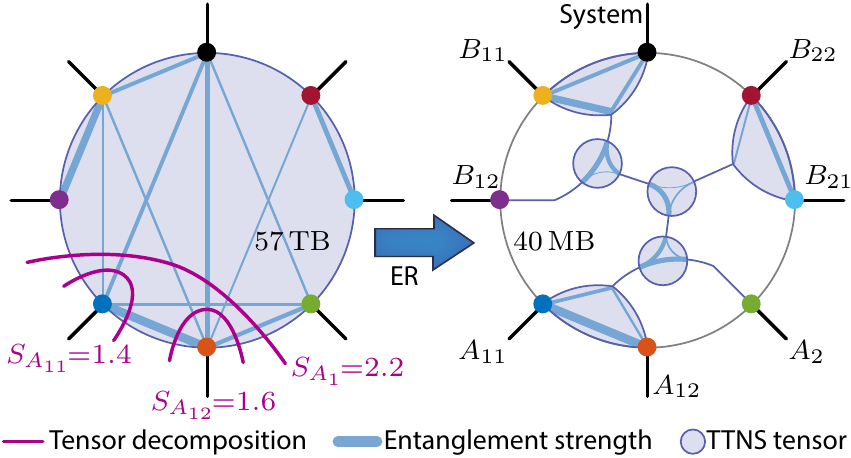}
    \caption{\textbf{Entanglement renormalisation for tensor compression.}
    	The 8th order tensor (left) is compressed from $\unit[57]{TB}$ down to $\unit[40]{MB}$ by replacement with a network of 3rd order ER-tensors (right).
        This network is designed to optimally encode the entanglement between all environments.
        Discrepancies in entropy $S$ between multiple tensor decompositions (purple) reveal the tensor's entanglement structure (blue) and guide the ER-network design.
	}
	\label{fig:er-network}
\end{figure}
%\textbf{We need a small bridge here...The essential idea underlying MPS and similar formats is that the vast majority of the exponentially large Hilbert space is never explored when the many-body entanglement generated by the dynamics remains low.
%Under this condition, the exponentially large tensor containing the amplitude of each many-body configuration of the system can be decomposed so that these amplitudes can be extracted by contracting  a product of much smaller objects (matrices or tensors)
%associated with each degree of freedom that can be stored and manipulated in a numerically tractable way to evolve the state. }

% curse of dimensions
Diagrammatically, the tensors (arrays) of a TTNS are represented as shapes with one open bond (line) per array dimension.
The multiplication of two tensors is drawn as a joining of the respective bonds (see Fig.~\ref{fig:er-network}).
A key figure regarding memory requirements is the bond dimension $D$, the size of the array dimensions and equal to the number of `auxiliary' states that encode the quantum correlations between neighbouring degrees of freedom  \cite{Schollwoeck2011,Orus2014,Vidal2007}.
$D$ thus governs the amount of entanglement that can be simulated, as measured by the von Neumann entropy $S=-\mathrm{Tr} (\rho \ln \rho)$, and typically ranges within $D \in [10,100]$ for acceptable results.
A single TTNS tensor scales exponentially $\mathcal{O}(D^N)$ in the number of nearest neighbours.
The chain tensors are relatively `cheap' to simulate (N=2), which is the essential motivation for the chain transformation.
However, the central tensor, representing the electronic system and connecting to each bath suffers a curse of dimensionality.
Although clustering and chain transformations reduced its $N$-neighbours from 252 in $\hat{\bm H}_{LVC}$ down to 7 in $\hat{\bm H}_{star}$, the central tensor still requires $\unit[57]{TB}$ of memory.

% ER as solution
Remarkably, we can compress the tensor down to $\unit[40]{MB}$ by decomposing it into a loop-free tree network of smaller auxiliary entanglement renormalisation (ER) tensors\cite{Vidal2007,Evenbly2009}(Fig.~\ref{fig:er-network}).
Their connectivity optimally exploits inter-environment entanglement (blue) to minimise the entropy $S$ across all bonds, which maximises the achievable accuracy at fixed $D$.
This ER-tree reduces complexity to $\mathcal{O}(ND^3)$ and is a key enabler for applying TTNS to $N>3$ environment models.

% TDVP for TTNS
Our TTNS is time-evolved with the time-dependent variational principle (TDVP)\cite{Haegeman2014,Lubich2014,Lubich2015,Schroder2016}, that has proven to be faster than competing algorithms\cite{Zwolak2004,Schroder2016}, and establishes intriguing links to MCTDHF \cite{Lubich2015}.
In this scheme, each tensor is sequentially time-evolved with a local effective Hamiltonian $\bm H^{eff}(t)$ that is constructed at each time step from the full many-body state, allowing long-ranged interactions, which is crucial for building and updating the ER-network.
All simulations are performed at zero temperature, with no environmental excitations, and start initially from the optically bright $\lep$.

\textbf{Fission dynamics in DP-Mes.}
\begin{figure*}
	\includegraphics[width=\linewidth]{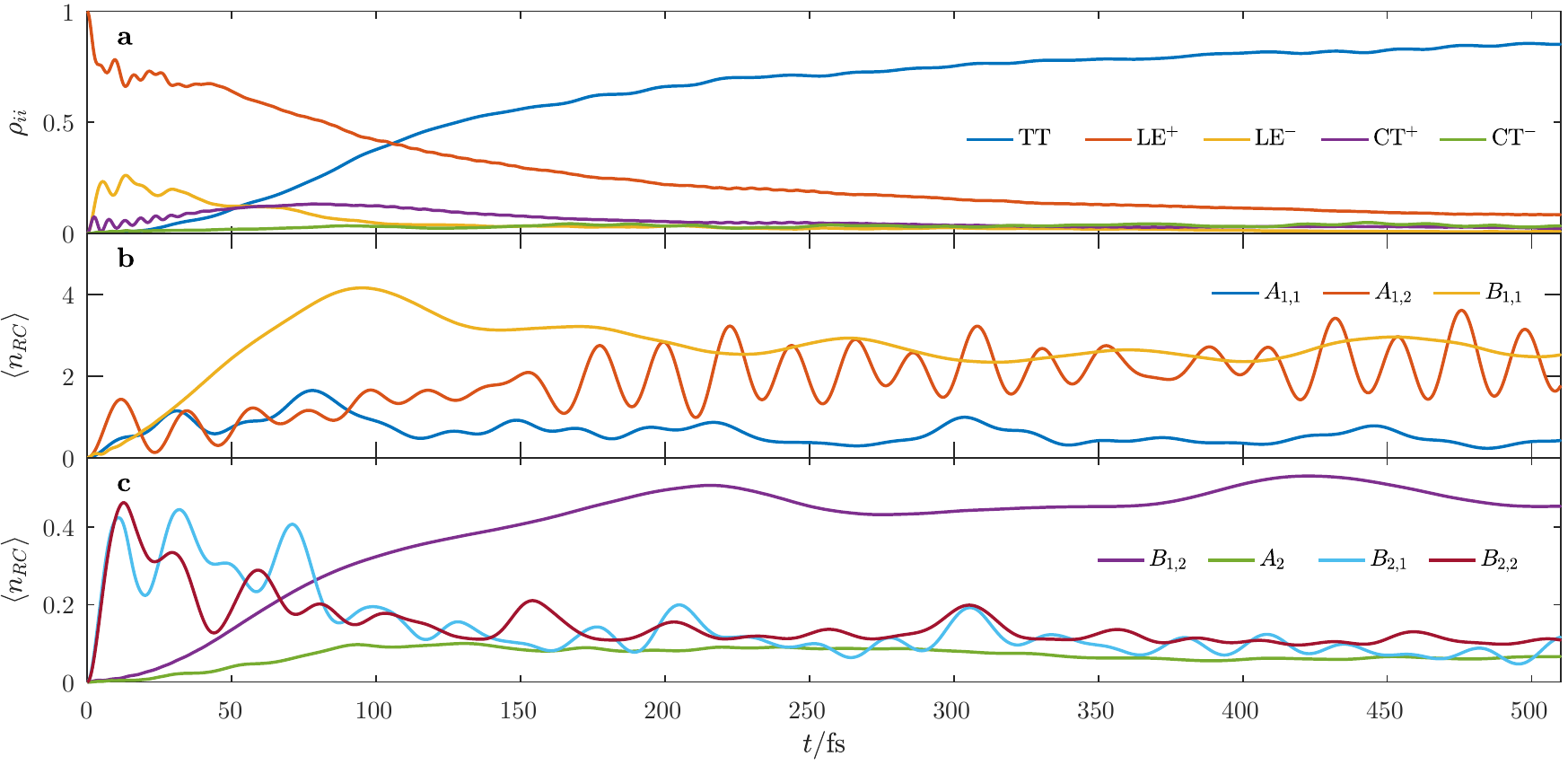}		% Dataset: 20170504-1825-10-DPMES-Tree4-Tree-v77TCMde9-L100CT0LE: 80-30
	\caption{\textbf{Simulated SF dynamics in DP-Mes.}
		Electronic populations (a) and environmental RC occupation dynamics (b,c).
		For $< \unit[20]{fs}$ damped Rabi oscillations transfer population from $\lep$ to $\lem$ and $\ctp$.
		At subsequent times TT mirrors the $\lep$ decay leading to SF on a timescale of $\unit[200]{fs}$ and suggesting a direct SF pathway.
		Similar timescales are present in the bath excitation $\braket{n_{RC}}$ of $B_1$ and $B_2$.
		The final TT yield reaches $90\%$ after $\unit[1]{ps}$.
	}
	\label{fig:pop-kin} 
\end{figure*}
% Fast multiscale SF
Figure \ref{fig:pop-kin} presents numerically exact electronic population dynamics (a) and the environmental occupation of the RCs across different orders of magnitude (b,c).
Vibrationally mediated SF happens on a timescale of $\unit[200]{fs}$ (Fig.~\ref{fig:pop-kin}c), with a final TT yield of $\approx90\%$, 
comparable to experiments where yields around $94\%$ and timescales of $\unit[400-700]{fs}$ have been reported\cite{Lukman2015}.
Considering the typical instrument response of $\unit[100]{fs}$ and the limitations of TD-DFT in vacuum, while experiments were performed in solutions, this result appears entirely reasonable.
% Rabi at early times
The dynamics are multiscale; at early times $< \unit[20]{fs}$ they are dominated by ultrafast damped vacuum Rabi oscillations ($\lep\xleftrightarrow{B_1}\ctp, \lep\xleftrightarrow{B_2}\lem$, see Fig.~\ref{fig:pop-kin}a),
with significant population transfer between $\lep$ and $\lem$ accompanied by rapid oscillatory excitation of the RCs of the $B_{2}$ environments coupling these states (see \sm{} Fig.~10). 
The Rabi oscillations of $\ctp$ correlate with the initial increase of TT population $d\rho_{\TT}/dt$ indicating a small contribution of path I through $A_2$ (Fig.~\ref{fig:rhoij-detailed}).
This oscillatory exchange of quanta results from nuclear motion that allows us to resolve the initial mixing of $\lepm$ and $\ctpm$ which allows super-exchange to occur \cite{Beljonne2013,Bakulin2015}.
% Superexchange & bath excitation
Following this mixing ($>\unit[50]{fs}$), the population of the TT state quickly begins to rise, mirroring the decay of the $\lep$ state, and accompanied by strong excitation of the $B_{1}$ RCs. 
These vibrations couple both $\lem$ and TT via $\ctm$, and thus open the vibronic super-exchange channel, labelled path II in Fig.~\ref{fig:states}.
The appearance of a strong imaginary coherence Im($\rho_{\TT,\lem}$) indicates coherent transport between respective states, which is only possible because the \emph{common} environment ($B_1$) mediates the coupling.
Since it also matches the major increase in TT population $d\rho_{\TT}/dt$, it is a strong evidence that the $B_1$ super-exchange drives SF (see Fig.~\ref{fig:rhoij-detailed}).
The final $5\%$ $\ctm$ content also shows that we reach a bound TT state which is lower in energy than a T+T (or non-relaxed TT state)\cite{Kolomeisky2014,Stern2017,Yong2017}.
The non-vanishing $\lepm$ on the other hand indicates that the equilibrium state can exhibit singlet character in its excited state absorption \cite{Lukman2015},
We also note that the TT population is modulated by the fast $A_{1,2}$ tuning modes which could be observable in ultrafast vibrational spectroscopy \cite{musser2015evidence}.
% Anharmonicity through coupling, leading to damping
Finally, although the model is purely harmonic, the vibrational motion we see is effectively damped.
This dissipation originates from anharmonicity induced by the tuning modes coupled to the electronic states, leading to energy transfer between coupling modes which can only occur through the presence and interaction of multiple independent environments.
Given the different roles of each environment in this system, an interesting idea to explore would be whether \emph{inter-environment} energy transfer could be further harnessed to control/coordinate optoelectronic processes.

% bound TT state, singlet in TT

\begin{figure}
	\includegraphics[width=\linewidth]{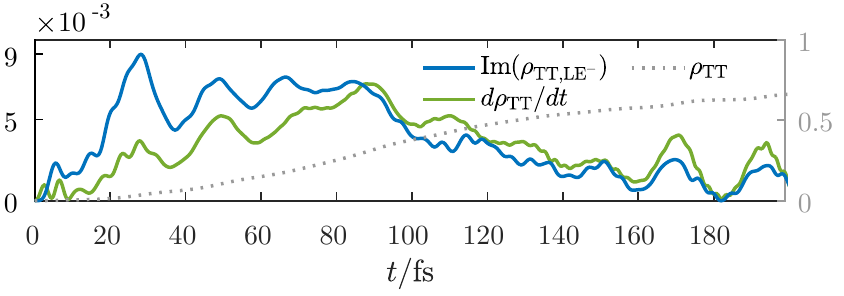} %20170409-1735-52-DPMES-Tree4-Tree-v77TCMde11-L100CT0LE
	\caption{\textbf{Evidence for coherent transport.}
		The change $d\rho_{\TT}/dt$ of $\rho_{\TT}$ correlates with the dynamic coherence Im($\rho_{\TT,\lem}$) indicating coherent super-exchange via $B_1$.
        The large oscillations with period of $\unit[20]{fs}$ correlate with $A_{1,2}$ oscillations (see Fig.~\ref{fig:pop-kin}b).
        The maximum SF rate is achieved at $\unit[90]{fs}$ during the passage of the avoided crossing in the TES.
	}
	\label{fig:rhoij-detailed}
\end{figure}

\textbf{Dynamic energy surfaces.}
\begin{figure*}
	\includegraphics[]{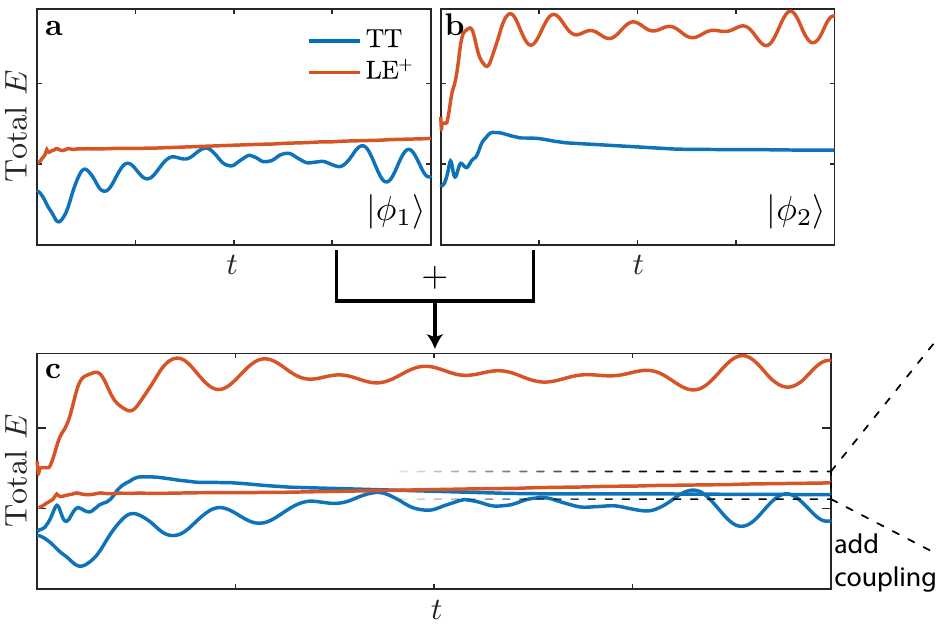}
	\includegraphics[]{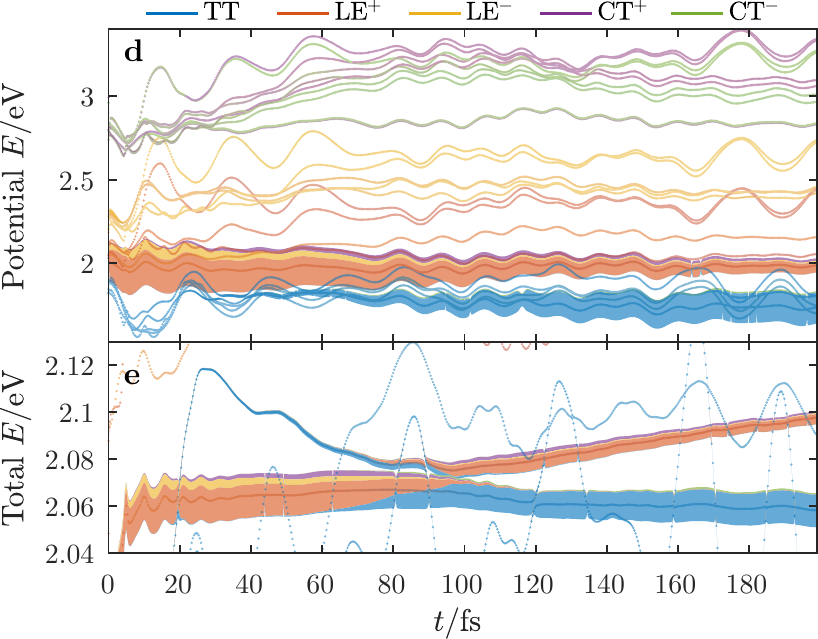} % 20170623-1055-33-DPMES-Tree4-Tree-v79-L100CT0LE+\results-Till1700Step0.5v79-OBBmax40-Dmax(80-30)-expvCustom256-1core-small.mat
	\caption{\textbf{Energy surfaces and their electronic populations.}
    	Each bath state $\ket{\phi_i}$ creates its own set of five energy surfaces (a,b,simplified).
        Combining them for five different $\ket{\phi_i}$, as sketched in (c), generates the presented 25 energy surfaces (d,e).
    	The filled areas indicate the amount of electronic population on each surface.
        The surfaces (dots) are coloured according to their mixing of electronic states.
        (d) The SF dynamics are not caused by conical intersections in the adiabatic PES.
        (e) Instead the vibronic mixing and super-exchange is sufficient to produce an avoided crossing at $\unit[80]{fs}$ of $\lep$ and TT in the non-adiabatic TES causing rapid SF.
	}
	\label{fig:pot-dyn} 
\end{figure*}
% Intro
In the following, we calculate adiabatic potential energy surfaces (PES) and total energy surfaces (TES), which also include the non-adiabatic nuclear kinetic energy operator
to gain a deeper understanding of the mechanism behind SF in DP-Mes.
% unravel multi-dim to time
A major difficulty of visualising energy surfaces for $N>2$ vibrational modes is that the coordinate $\bm x = (x_1,\ldots,x_N)$ is multi-dimensional.
Usually, this problem is overcome by presenting the potentials along two selected modes which are considered relevant reaction coordinates.
We avoid this limitation by plotting the surfaces against time $t$, to obtain reaction energy profiles.
These follow the optimal paths chosen by the exact wave function dynamics, unravelling them into a single dimension (see Fig.~\ref{fig:pot-dyn}).

% How to obtain
We calculate the energy surfaces on-the-fly as the eigenvalues of the effective Hamiltonian $\bm H^{eff}(t)$ and potential $\bm V^{eff}(t)$
\begin{align}
	H^{eff}_{(ij),(k,l)}(t) &= \bra{\phi_i(t)}\bra{j}\hat{\bm H}\ket{\phi_k(t)}\ket{l},\\
    V^{eff}_{(ij),(kl)}(t) &= \bra{\phi_i(t)}\bra{j}\hat{\bm H}-\hat{\bm T}\ket{\phi_k(t)}\ket{l},
\end{align}
where $i,k$ index bath states and $j,l$ denote the electronic states.
We simultaneously consider five different bath states $\ket{\phi_i(t)}$, where each $\ket{\phi_i(t)}$ relates to a many-body superposition of elementary, 
localised phonon wave packets $\ket{\varphi_j(\bm x(t))}$ (see \sm{} Sec.~V) and are mutually orthogonal, i.e. $\braket{\phi_i(t)|\phi_j(t)}=\delta_{ij}$. 
These states can be found at each time point from the singular value decomposition (SVD) of the total (pure) wave function for a bipartition into system and environment degrees of freedom. 
The SVD confirms that there are at most five environmental configurations, which generate an effective Hamiltonian $H_{eff}(t)\in \mathbb{C}^{25\times25}$, leading to 25 surfaces (Fig.~\ref{fig:pot-dyn}a-c, simplified for two states and two configurations). 
Indeed, the SVD is a central element in the numerical algorithm for updating and normalising the TTNS, enabling facile extraction of these configurations.
Our method is basis-free, independent on the choice of local oscillator wave functions, since we use $\bm V^{eff}(t)$ and $\bm H^{eff}(t)$ only.
Here, the wave function mostly populates only two surfaces (see Fig.~\ref{fig:pot-dyn}d,e) which we will focus on in the following.
% basis free
% What we are looking at now
In contrast to the population and occupation dynamics we now look into the energetics of the dynamics (Fig.~\ref{fig:pot-dyn}d,e)
which are revealed as changes of the energy level (dots), the overall population of each surface (fills), and even the mixing of diabatic electronic states (colour of fills) on the surfaces.

% General picture
For each electronic state we can observe five energy surfaces (Fig.~\ref{fig:pot-dyn}d), each caused by a different nuclear configuration $\ket{\phi_i(t)}$.
Oscillatory features are created by normal mode displacements and can be assigned to specific environments.
As an example, the dominant oscillations with $\approx \unit[20]{fs}$ period visible in the highest PES of $\ctpm$ are created by $A_{1,2}$ displacements tuning the energy levels (cf. Fig.~\ref{fig:pop-kin}b).
Avoided crossings between energy surfaces indicate coupling between the respective diabatic states, accompanied by state mixing expressed in changing surface colour (see Fig.~\ref{fig:pot-dyn}e).
The SF dynamics in the adiabatic PES (Fig.~\ref{fig:pot-dyn}d) are evident as population transfer from the lowest $\lep$ surface to the lowest TT surface.
Since the population transfers between non-crossing surfaces, this displays the non-adiabatic nature of the dynamics.

% B22 driving initial TT rise, No Conical Intersection, avoided crossing
The adiabatic PES of the populated TT (blue) stays consistently below the $\lep$ PES (red, Fig.~\ref{fig:pot-dyn}d), thus SF in DP-Mes should not be caused by a conical intersection as proposed for other SF systems\cite{musser2015evidence,Miyata2017}.
The crossings of the highest energy TT PES and the main $\lep$ PES are not conical intersections and do not transfer populations, as they correspond to spatially separated and uncorrelated environmental states.
Instead, SF is driven by an avoided crossing visible in the non-adiabatic TES (Fig.~\ref{fig:pot-dyn}e) at $\unit[80]{fs}$, caused by three main factors.
First, the unoccupied TT TES is pushed above the $\lep$ TES at $\unit[20]{fs}$ by $A_{12}$ and $B_{22}$, as indicated by correlations with their RC displacement and excitation (see \sm{} Fig.~7).
Secondly, the TT is effectively coupled to the $\lep$ by super-exchange, resulting in the avoided crossing gap of $\unit[11]{meV}$.
The third factor is the lowering of TT through the avoided crossing which leads to population transfer onto TT.

% RC frequencies -> coupling & lowering
The above time scales are found in the environment's RC frequencies, reflecting properties of the collective motion, giving us more information about the underlying mechanisms.
While the energy increase of the (unoccupied) TT TES at $\unit[20]{fs}$ is caused by the response of the collective modes $\omega_{RC,A_{12}} \approx \unit[1377]{cm^{-1}} \sim \unit[24]{fs}$ and $\omega_{RC,B_2} \approx \unit[1100]{cm^{-1}} \sim \unit[30]{fs}$ to the sudden occupation of the $\lep$ state, the effective vibronic coupling emerges from displacement of $B_2$ and $\omega_{B_1} \approx \unit[400]{cm^{-1}} \sim \unit[83]{fs}$, setting a longer time scale for SF (Fig.~\ref{fig:pop-kin}b,d).
The slow reorganisation of $B_1$ increases the effective TT-$\lep$ coupling from $\unit[3]{meV}$ at $\unit[20]{fs}$ to $\unit[6]{meV}$ at $\unit[80]{fs}$, as seen from the avoided crossing gaps.
Crucially, the $B_1$ modes also lead to the TT TES descent, slow enough in the Landau-Zener sense to minimise the residual $\lep$ population on the upper surface, in contrast to the rapid crossing at $\unit[20]{fs}$,
where the effective coupling is weaker and no population transferred.
Estimates based on the Landau-Zener formula for the diabatic transition probability predict a coupling of $\unit[10]{meV}$ which is encouragingly close to our result (see \sm{} Sec.~VI~E).
This confirms that vibrationally mediated SF can be fast without the need for a conical intersection.
Moreover, it suggests that SF rates and yields might be controlled by altering RC vibrational frequencies by isotopic substitutions, for example. 
% Not A2
Finally, the more direct path I via $A_2$ has only minor contribution to SF, since its TT projected RC occupation rapidly drops, which can only happen if TT was populated through $\ctm$ (see \sm{} Sec.~VI~D).

\textbf{Conclusions.} %TODO: Sell it better!
% All brand-new techniques developed
By uniquely combining a number of recent approaches to many-body simulation, we have shown how the non-perturbative dynamics of a realistically parameterised molecular dimer can be obtained and microscopically analysed by tensor network and machine learning methods.
Understanding the entanglement `topology' of many-body entangled states is essential for an appropriate entanglement encoding and state compression, which is at the heart of TTNS, as well as neural network state\cite{Carleo2017} approaches.
The crucial advance is the introduction of a flexible network topology that optimally distributes and encodes the strong entanglement between the core system and its multiple environments into numerically tractable elements.
This understanding is provided by multi-environment open system models as general framework and can be transferred to more complex applications in strongly correlated electrons in quantum chemistry \cite{Szalay2015} and correlated materials design.
The insights possible from exploiting TTNS-like methods are highlighted in our SF example, where we have identified the underlying mechanism, origin of timescales and, hence, the sequence of multi-environment reorganisations that drive efficient SF in a symmetry forbidden dimer.
Further investigation of the real-time role of vibronic effects and dynamical symmetry breaking may suggest other ways to optimise SF, such as reducing triplet-triplet annihilation or promoting their spin-decorrelation and spatial separation.
Indeed, the more general idea of designing functional molecular and condensed matter nanostructures optimised for multiscale non-equilibrium properties is an intriguing, albeit challenging idea \cite{bredas2017photovoltaic,romero2017quantum,scholes2017using},
for which powerful methods for simulating \textit {ab inito}-parameterised open quantum dynamics will be essential.

\section*{Methods}
\subsection*{DFT}

All DFT calculations were performed with the NWChem electronic structure code in vacuum \cite{nwchem}.
The ground-state structure and vibrational modes were obtained employing analytical Hessians at the cc-PVDZ/B3LYP
level of theory. 
Excited state energies and forces (corresponding to the diagonal elements of $\bm W$) for $\text{LE}^{\pm}$ 
and $\text{CT}^{\pm}$ were calculated from (linear-response) TDDFT gradients at the cc-PVDZ/LC-BLYP level of theory. 
The long-range corrected functional is required to correctly describe the $S_1$ state of pentacene as well
as the states with charge-transfer character. An optimised range-separation parameter $\mu=0.29$ was used 
in the LC-BLYP functional. This choice gives a good description of the energy of the first excited singlet of
the pentacene molecule \cite{Wong2010}.
For TT the quintet state (total spin 2) was used as a
proxy for the purpose of calculating energy and forces.
The frontier molecular orbitals at the cc-PVDZ/LC-BLYP level were also used as inputs for the evaluation of the off-diagonal couplings.
Modes below $\unit[110]{cm^{-1}}$ and above $\unit[1500]{cm^{-1}}$ were disregarded due to either unrealistic \textit{ ab initio} parameters arising from the neglect of their anharmonicities or irrelevance for present time-resolved experiments, respectively.
Limitations of the method are discussed in the \sm{} Sec.~I.

\subsection*{Tensor Network States}
To prepare the linear vibronic DP-Mes Hamiltonian for the TTNS simulations, the vibrational modes were clustered to independent environments by using a weighted k-means algorithm\cite{Arthur2007}.
These environments were then mapped onto chain Hamiltonians with the orthogonal polynomials transformation to facilitate the tensor tree network decomposition and numerical evolution of the complete vibronic wave function \cite{Chin2010} 

The model was simulated with a TTNS as depicted in \sm{} Fig.~4, incorporating entanglement renormalisation (ER) tensors connecting the vibrational chain environments to the excitonic states 
and using an optimised boson basis (OBB) to allow for a large and expandable Fock basis.
The time-evolution is performed in the time-dependent variational principle (TDVP)\cite{Haegeman2014}
to simulate exciton-phonon dynamics at zero temperature with no initial environmental excitations at a time step of $\unit[0.33]{fs}$.
Details about the TDVP algorithm we have developed for the TTNS evolution are given in the \sm{} Sec.~II and further background can be found in Refs.~\onlinecite{Haegeman2014, Lubich2014, Szalay2015, Schroder2016, Shi2006}.
% Maximum Bond dimensions
We keep 100 bosonic Fock states per vibrational mode.
The dynamics converged sufficiently for an OBB dimension of 40, maximum TTNS bond dimension $D$ between ER-nodes of $D_{Node,max} = 80$, while along the chains $D_{Chain,max}= 30$ was appropriate (see \sm{} Fig.~5).
Thus we cover a Hilbert space of $10^{500}$ states using only $10^7$ parameters.
Energy surfaces were calculated from the effective Hamiltonian and effective potential of the system tensor.
Further computational details are provided in \sm{} Sections~III, IV, and V.

\subsection*{Data Availability}
The data underlying this publication are available in the University of Cambridge data repository at URL. % TODO

\subsection*{Code Availability}
The custom tree tensor networks state code used for simulations can be accessed at URL. % TODO

\bibliography{DPMES}

\begin{acknowledgments}
\noindent We thank R.H.~Friend for making this work possible. 
F.A.Y.N.S., D.H.P.T., N.D.M.H., and A.W.C. gratefully acknowledge the support of the Winton Programme for the Physics of Sustainability and the Engineering and Physical Sciences Research Council (EPSRC).
\end{acknowledgments}

\section*{Author Contributions}
F.A.Y.N.S. developed the TTNS method and performed calculations and D.H.P.T. performed DFT calculations. 
F.A.Y.N.S., D.H.P.T. and A.W.C. interpreted results and wrote the manuscript. 
A.J.M. initiated the project and provided experimental expertise. 
A.J.M., N.D.M.H. and A.W.C. supervised the project. 
All authors discussed results and contributed to the manuscript.

\section*{Competing financial interests}
The authors declare no competing financial interests.

%\clearpage
%\setcounter{section}{0}
%\setcounter{figure}{0}
%
%
%\include{SupplementalContent}

\end{document}